\documentclass[submission,copyright,creativecommons]{eptcs}
\providecommand{\event}{PROLE 2014} 

\usepackage[T1]{fontenc}
\usepackage[latin1]{inputenc}
\usepackage{graphicx}
\usepackage{subfig}

\usepackage{times}
\usepackage{latexsym}
\usepackage{amssymb}
\usepackage{stmaryrd}
\usepackage{epsf}
\usepackage[arrow,matrix,tips]{xy}

\newtheorem{definition}{Definition}[section]

\title{SpecSatisfiabilityTool: 
A tool for testing the satisfiability of specifications on XML documents}
\author{
Javier Albors \qquad\qquad Marisa Navarro
\institute{Departamento de LSI, UPV/EHU\\
San Sebasti\'an, Spain}
\email{\quad jalbors001@gmail.com \quad\qquad marisa.navarro@ehu.es}
}
\def\titlerunning{SpecSatisfiabilityTool}
\def\authorrunning{J. Albors,  M. Navarro}
\begin{document}
\maketitle

\begin{abstract}
We present a prototype that implements a set of logical rules to prove the satisfiability for a class of specifications on XML documents. 
Specifications are given by means of constrains built on Boolean XPath patterns. 
The main goal of this tool is to test whether a given specification is satisfiable or not, and justify the decision showing the execution history. 
It can also be used to test whether a given document is a model of a given specification and, as a by-product, 
it permits to look for all the relations (monomorphisms) between two patterns and to combine patterns in different ways.
The results of these operations are visually shown and therefore the tool makes these operations more understandable.
The implementation of the algorithm has been written in Prolog but the prototype has a Java interface for an easy and friendly use. 
In this paper we show how to use this interface in order to test all the desired properties.
\end{abstract}

\section{Introduction}

Our aim is to define specifications of XML documents as sets of constraints (of some specific class) on these documents, and to provide a form of reasoning about these specifications. 
XML documents will be represented by trees and the constraints will be based on some kind of XPath queries \cite{WWWC2007, BK08, BFG08}.

To define the constraints on some XPath notation, we have selected the representation of Boolean XPath queries given in \cite{MS04}, where Miklau and Suciu study the containment and equivalence problems for a class of XPath queries that contain branching and label wildcards and can express descendant relationships between nodes. In particular, they introduce \emph{Boolean patterns} as an alternative representation of this class of queries. These patterns are trees consisting of nodes with labels (or $*$, for a wildcard in the query) and two kinds of edges, child edges (/) and descendant edges (//),  for the corresponding axes in the query.  For instance, the pattern $p$ in Figure~\ref{fig:monomorphism}  (on the left)  corresponds to the XPath expression $/a[b][.//*[e][d]]$. 
We define three sorts of constraints (positive, negative, and conditional constraints)  on these patterns. A specification is defined as a set of clauses, where a clause is a disjunction of constraints. 

Our main question is about satisfiability, that is, given a specification $\mathcal{S}$, whether or not there exists an XML document satisfying all constraints in $\mathcal{S}$. Moreover, we are looking for adequate inference rules to build a sound and complete refutation procedure for checking satisfiability of a given specification. In addition to checking satisfiability, these rules would be used to deduce other constraints, which can permit us to optimize a satisfiable specification.

Our approach follows the main ideas given in \cite{OEP08} (here it is shown how to use graph constraints as a specification formalism on graphs and how to reason about these specifications, providing refutation procedures based on inference rules that are sound and complete) and try to apply such ideas to XML documents. 
However, the particularization of graph constraints to our setting is not trivial (mainly because our patterns are more expressive). Similarly, our inference rules take a similar format to the inference rules given in \cite{OEP08}, but the particularization to our setting needs to define appropriate operators and to prove new results.
The formal study of our work is now submitted for presentation, but the ideas and preliminaries of such work were introduced in \cite{NO10}. 

In this paper we focus on the prototype that implements our refutation procedure. The algorithm is written in Prolog \cite{CM03} but also has a Java interface for an easy and friendly use. 
The main goal of this tool is to test whether a given specification is satisfiable or not, and justify the decision by showing the rules applied during the procedure execution.
It can also be used to test whether a given document is a model of a given specification and, as a by-product, it permits to look for all the monomorphisms between two patterns or to look for the result of doing the operations  $p\otimes q$ and $p\otimes_{c,m}q$ which are necessary for implementing some rules. The results of these operations are visually shown and therefore it makes them more understandable.

The paper is organized as follows. There are two main sections, the first one dedicated to the formal background and the second one to show the prototype. 
Section \ref{fb} starts by introducing the formal definitions of $document$, $pattern$, and the relations (monomorphisms) between them. 
Then, in Section \ref{espec}, we present the constraints and clauses that we are going to use to define our specifications and, 
in the Section \ref{rules}, the main inference rules for our refutation procedure. 
The prototype is presented in Section \ref{prot} where we explain how to perform several operations by means of examples and by showing different screenshots of the tool.
Finally, in Section \ref {imp notes}, we give some notes on implementation and some comments about the ongoing work for obtaining completeness for our refutation procedure.  

%

\section{Formal Background} \label{fb}

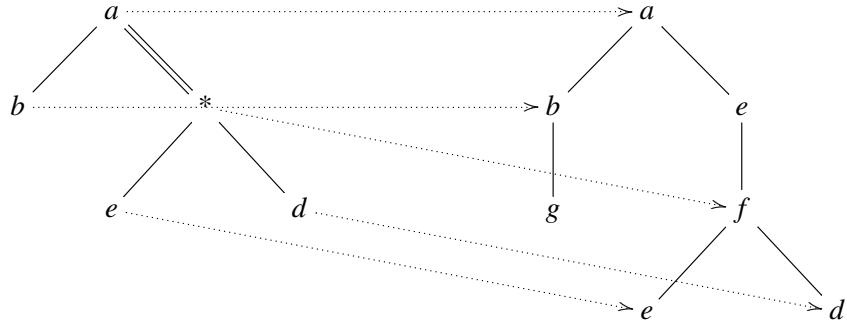
\begin{figure} 
$$
    \xymatrix{
          & a  \ar@{-}[dl] \ar@{=}[dr] \ar@{.>}[rrrrrr] &&&&&
          & a \ar@{-}[dl] \ar@{-}[dr] \\ 
          b \ar@{.>}[rrrrrr] && \textrm{*} \ar@{-}[dr] \ar@{-}[dl] \ar@{.>}[drrrrrr] &&&&
          b \ar@{-}[d] && e \ar@{-}[d] \\
         & e \ar@{.>}[drrrrrr] && d \ar@{.>}[drrrrrr] &&&
         g && f  \ar@{-}[dl] \ar@{-}[dr] \\
        &&&&&&& e && d\\ }
$$
\caption{A monomorphism $h: p\rightarrow t$ from a pattern $p$ to a document $t$}
\label{fig:monomorphism}
\end{figure}

We consider an \emph{XML document} as an unordered and unranked tree with nodes labelled from an infinite alphabet $\Sigma$. 
The symbols in $\Sigma$ can represent the element labels, attribute labels, and text values that can occur in XML documents. 
Note that this is an abstract representation of a real XML document since we are only interested in its tree structure. 
Figure~\ref{fig:monomorphism} shows a document $t$ (on the right) with root labelled $a$ and two subtrees.
Here is the formal definition of document.

\begin{definition}
Given a signature $\Sigma$, a \emph{document on $\Sigma$} is a tree $t$ whose nodes are labelled with symbols from $\Sigma$ and with one sort of edges denoted /. 
$Nodes(t)$ and $Edges(t)$ denote respectively the sets of  nodes and edges in $t$; $Root(t)$ denotes its root node; and for each $n\in Nodes(t)$, $Label(n)$ denotes the label of such a node $n$. Each edge in $Edge(t)$ is represented $(x,y)$ with $x, y\in Nodes(t)$. Each $(x,y)\in Edges^+(t)$ represents a path in $t$ from node $x$ to node $y$.
\end{definition} 

As said in the introduction, we use patterns as an alternative representation of Boolean queries. 
Patterns will be also represented as some sort of trees but with two differences with respect to documents. Some edges in patterns can be double (//) and it is permitted the label $*$ in nodes. 
Figure~\ref{fig:monomorphism} shows a pattern $p$ (on the left) that has a label $*$  in a node and an edge // from the root $a$ into one of its children.

A pattern specifies the conditions that a document must hold. For instance, the pattern $p$ in Figure~\ref{fig:monomorphism} specifies the following conditions: ``The root of the  document must be a node labelled $a$, one of its child nodes must be a node labelled $b$, and a descendant node of the root  must have (at least) two children labelled $e$ and $d$''.
Here is the formal definition of pattern.

\begin{definition}
Given a signature $\Sigma$, a \emph{pattern on $\Sigma$} is a tree $p$ whose nodes are labelled with symbols from $\Sigma\cup\{*\}$ and with two sorts of edges: 
the descendant edges denoted //, and the child edges denoted /. $Nodes(p)$, $Edges(p)$, $Root(p)$, and $Label(n)$ are defined as in the previous definition; but now the edges are distinguished: $Edges(p)$ = $Edges_{//}(p)\cup Edges_{/}(p)$, therefore $(x,y)\in Edges^+(p)$ represents a path in $p$ from node $x$ to node $y$ with edges of type / or //  along the path.
\end{definition}

To define when a document satisfies a given pattern we use the notion of homomorphism. As documents are patterns without labels $*$ or edges //, 
we define here the notion of homomorphism between two patterns and, as a consequence, the definition of homomorphism from a pattern into a document is a particular case.

\begin{definition} \label{embedding-def}
Given two patterns $p$ and $q$, a \emph{homomorphism} from $p$ into $q$  is a function 
$h: Nodes(p) \rightarrow Nodes(q)$ satisfying the following conditions:
\begin{itemize}
\item Root-preserving: $h(Root(p))=Root(q)$;
\item Label-preserving: For each $n\in Nodes(p)$, $Label(n)=*$ or 
   $Label(n)=Label(h(n))$;
\item Child-edge-preserving: For each $(x,y)\in Edges_{/}(p)$,  $(h(x),h(y))\in Edges_{/}(q)$.
\item Descendant-edge-preserving: For each $(x,y)\in Edges_{//}(p)$, 
   $(h(x),h(y))\in Edges^{+}(q)$.
\end{itemize}
\end{definition}

Note that the ``child-edge-preserving'' condition says that each edge /  in a pattern $p$ must be applied into an edge / in the pattern $q$ and the ``descendant-edge-preserving'' condition says that each edge //  in a pattern $p$ can be applied into a path in $q$ (with one or more edges of type / or //  along the path).

In the particular case when $q$ is a document, the last condition applies each edge //  in the pattern $p$ into a path in the document $q$ (with one or more edges / along the path).

\begin{definition} 
Given a pattern $p$ and a document $t$, we say that \emph{t satisfies p}, denoted $t \vDash p$, if there exists a monomorphism (i.e, an injective homomorphism) from $p$ into $t$. 
The model set of a pattern $p$ is the set of documents satisfying $p$: $Mod(p) = \{ t \;\mid \;   t \vDash p\}$.
\end{definition}

In Figure~\ref{fig:monomorphism} there is a pattern $p$ (on the left), a document $t$ (on the right) and a monomorphism $h: p\rightarrow t$ (which is drawn with dotted arrows). 
Then $t$ satisfies (or is a model of) $p$. We can see that in fact the root of the document $t$ is a node labelled $a$, one of its child nodes is a node labelled $b$, and a descendant node of the root  (in this document the node labelled $f$) has two children labelled $e$ and $d$. 
It corresponds to evaluate the XPath expression $/=/a[b][.//*[e][d]]$ against the document $t$.

\subsection{Specifications}\label{espec}
 
We assume that a specification consists of a set (or conjunction) of clauses, where a clause is a disjunction of constraints (the empty disjunction is the clause \emph{FALSE}).
Now we introduce the three kinds of constraints we are going to use: positive, negative, and conditional constraints. 
A positive constraint specifies that a pattern must be satisfied and a negative constraint specifies that the pattern must not be satisfied. 
A conditional constraint consists of two patterns, $p$ and $q$, such that $q$ is an extension of $p$. 
Roughly speaking, this constraint specifies that whenever a document $t$ verifies the pattern $p$ it should also verify the extended pattern $q$ (see Definition \ref{defSat}).

\begin{definition}
Given two patterns $p$ and $q$, we say that $p$ is a {\em prefix} of $q$ if there exists an injective function (called {\em prefix functio}n)
$c: Nodes(p) \rightarrow Nodes(q)$  satisfying the following conditions:
\begin{itemize}
\item Root-identity: $c(Root(p))=Root(q)$;
\item Label-identity: For each $n\in Nodes(p)$,  $Label(n)=Label(c(n))$;
\item Child-edge-identity: For each $(x,y)\in Edges_{/}(p)$, $(c(x),c(y))\in Edges_{/}(q)$;
\item Descendent-edge-identity: For each $(x,y)\in Edges_{//}(p)$, $(c(x),c(y))\in Edges_{//}(q)$.
\end{itemize}
\end{definition}

\begin{definition}
Given a pattern $p$,  $\exists p$ denotes a {\em positive constraint} and  $\neg\exists p$ denotes a {\em negative constraint}.
A {\em conditional constraint} is denoted $\forall(c: p  \rightarrow q)$ 
where $p$ and $q$ are patterns, $p$ is a prefix of $q$ with $c: Nodes(p)  \rightarrow Nodes(q)$ being the prefix function. 
\end{definition}

The satisfaction of clauses is defined inductively as follows.

\begin{definition}\label{defSat}
A  document $t$ satisfies a clause $\alpha$, denoted $t \models \alpha$, if it holds:
\begin{itemize}
\item $t \models \exists p$ if  $t \vDash p$  (that is, if there exists a monomorphism $h: p  \rightarrow t$);
\item $t \models \neg \exists p$ if  $t \nvDash p$ (that is, if there does not exist a monomorphism $h: p  \rightarrow t$);
\item $t \models \forall(c: p  \rightarrow q)$ if for every monom. $h: p  \rightarrow t$ 
          there is a monom. $f: q  \rightarrow t$ such that $h = f \circ c$.
\item $t \models L_1 \vee L_2\vee \dots \vee L_n$ if 
          $t \models L_i$ for some $i\in \{1,\dots ,n\}$.
\end{itemize}
\end{definition}

Let us see now an example of a conditional pattern.  Let $p$ be the pattern corresponding to the XPath expression $/a[.//e]$ (that is, the tree root is labelled $a$ and the child node is labelled $e$ with an edge // between both nodes) and let $q$ be the pattern corresponding to the XPath expression $/a[.//e[f]]$ (that is, the tree extending $p$ by adding a node labelled $f$ as a child node of $e$ with an edge / between them). 
The document $t$ in Figure~\ref{fig:monomorphism} (on the right) does not satisfy the constraint $ \forall(c: p  \rightarrow q)$ 
(where $c$ is the prefix function applying $p$ into $q$) 
since not all descendant nodes labelled $e$ in $t$ have a child labelled $f$. However the  document $t$ satisfies the pattern $q$. 
Note that  in general to verify the conditional constraint  $ \forall(c: p \rightarrow q) $ is stronger than to verify the clause  $C$ = $\neg\exists p \vee\exists q $.


\subsection{Inference Rules for a Refutation Procedure} \label{rules}

 A {\em refutation procedure} for a specification
$\mathcal{S}$ can be seen as a sequence of inferences
$\mathcal{C}_0$ $\Rightarrow$ $\mathcal{C}_1 \Rightarrow \dots$
$\Rightarrow \mathcal{C}_i \Rightarrow
\dots$ where the initial state is the original specification (i.e., $\mathcal{C}_{0} = \mathcal{S}$)
and each $\mathcal{C}_{i+1}$ is obtained from 
$\mathcal{C}_{i}$ by applying a rule.
The main inference rules of our refutation procedure are the following:
$$
\fbox{\begin{minipage}{3in}
$$\frac{\exists  p_1 \vee \Gamma_1 ~ ~ ~ ~ \neg \exists p_2 \vee \Gamma_2} 
{~ ~ \Gamma_1 \vee \Gamma_2} {\quad \textbf{ (R1)} }$$ \textrm{~ ~if there exists a monomorphism} $m: p_2 \rightarrow p_1$
\end{minipage}}
\vspace{0.2cm}
$$
Rule (R1) is like a resolution rule, since the two premises have literals that are, in some sense, ``complementary'': one is a positive constraint, the other one is a negative one, and 
the condition about the monomorphism from $p_2$ to $p_1$ plays the same role as unification. Note that when $\Gamma_1$ and $\Gamma_2$ 
are empty, the rule (R1) infers the clause \emph{FALSE}.
$$
\fbox{\begin{minipage}{3in}
$$\frac{\exists  p_1 \vee \Gamma_1 ~ ~ ~ ~ ~ \exists p_2 \vee \Gamma_2}
{~ (\bigvee_{s \in {p_1 \otimes p_2}} \exists  s) \vee \Gamma_1 \vee \Gamma_2} {\quad
\textbf{ (R2)}}$$
\end{minipage}}
\vspace{0.2cm}
$$
Rule (R2) builds a disjunction of positive constraints from two positive constraints. It uses the operator $\otimes$ that we define below. Informally speaking, $p_1$$\otimes$$p_2$ denotes the set of patterns that can be obtained by ``combining'' $p_1$ and $p_2$ in all possible ways.
$$
\fbox{\begin{minipage}{5,9in}
$$\frac{\exists  p_1 \vee \Gamma_1 ~ ~ ~ ~ \forall (c: p_2 \rightarrow q)\vee \Gamma_2}
{~ (\bigvee_{s \in p_1 \otimes_{c,m} q} \exists  s) \vee \Gamma_1 \vee \Gamma_2} {\quad\textbf{ (R3)} }$$ 
\textrm{if there is a monomorphism} $m: p_2 \rightarrow p_1$
\textrm{that cannot be extended to} 
$f: q\rightarrow p_1$ 
\textrm{such that} $f\circ c = m$.
\end{minipage}}
\vspace{0.2cm}
$$
Rule (R3) is similar to rule (R2): From a positive constraint $\exists p_1$ and a conditional constraint $\forall (c: p_2\rightarrow q)$, it builds a
disjunction of positive constraints. It uses the operator $\otimes_{c,m}$ that we define below. 
Informally speaking, 
$p_1\otimes_{c,m}q$ denotes the set of patterns that can be obtained by combining $p_1$ and $q$ in all possible ways, but maintaining $p_2$ shared. 

\begin{definition}
Given two patterns $p_1$ and $p_2$, $p_1\otimes p_2$ is the following set of patterns: 
$p_1\otimes p_2$ = \{$s$ $\mid$ there exist jointly surjective monomorphisms 
$inc_1: p_1\rightarrow s$ and $inc_2: p_2\rightarrow s$\} .
\end{definition}

\begin{definition}
Given two patterns $p_1$, $p_2$, a prefix function 
$c: p_2 \rightarrow q$, and a
monomorphism $m: p_2 \rightarrow p_1$,  
$p_1\otimes_{c,m}q$ is the following set of patterns:
$p_1\otimes_{c,m}q$ = \{$s$ $\mid$ there exist jointly surjective monomorphisms $inc_1: p_1\rightarrow s$ and $inc_2: q\rightarrow s$ such that $inc_1\circ m$ = $inc_2\circ c$\}.
\end{definition}

We have formally proven that the refutation procedure consisting of the three inference rules (R1), (R2), and (R3) is sound \cite{NO10}.
That is, whenever the procedure infers the clause \emph{FALSE} from a input set of clauses $\mathcal{S}$, then $\mathcal{S}$ is unsatisfiable. 
The prototype we explain in the next section implements this refutation procedure when we choose to execute ``Version 1''. 
Moreover, the refutation procedure also uses sound rules
for deleting and simplifying clauses 
and the implementation applies them as soon as possible to get a better performance. 
To sum up, given a specification as input, if the result of running ``Version 1'' is that the procedure stops with \emph{FALSE}, then we are sure that the specification is unsatisfiable.

However, the procedure is not complete: It may happen that the clause \emph{FALSE} is not inferred although $\mathcal{S}$ is unsatisfiable (see  \cite{NO10}).
Looking for a complete procedure, we have studied how to transform a positive constraint containing a descendant edge (//) into a (semantically equivalent) disjunction of positive constraints, in order to apply inference rules that could not be applied before such transformation. We call it  ``the unfolding process'' and have incorporated it into the refutation procedure. 
Some preliminary details of this study  can be found in \cite{Albors2014} and a preliminary refutation procedure obtained by adding this ``unfolding process'' has been implemented and can be tested running ``Version 2'' of the prototype. 
Although we are still working on a formal proof, we believe that the new procedure is complete. This would mean that if the input specification is unsatisfiable then the procedure stops and returns \emph{FALSE}.

Finally, we must observe that for satisfiable specifications, the procedure can stop (without obtaining the clause \emph{FALSE}) or not stop. We are studying the causes of non-termination and we would like to obtain that our procedure does not stop only in the case of satisfiable specifications whose models are all infinite. Such specifications are possible due to the conditional constraints. If we restrict to specifications with only positive and negative constraints, the refutation procedure is finite. 

\section{Showing the Prototype} \label{prot}

In this section we explain how to use the application. 
In particular, introducing the clauses, executing the refutation procedure, testing whether a document is a model of a specification, and other operations that can be visually executed.

\begin{figure}[h]
  \centering
    \includegraphics[width=\textwidth]{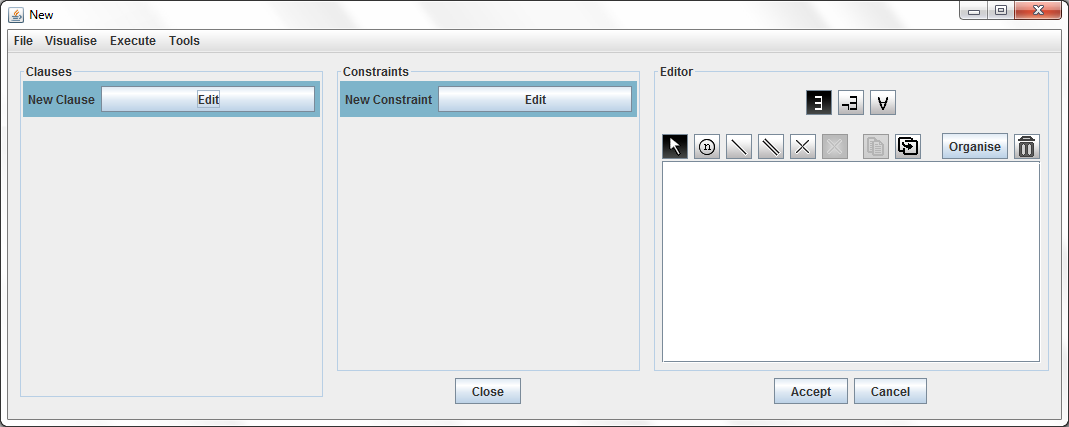}
  \caption{Home screen of the application}
  \label{fig:grafico1}
\end{figure}
\begin{figure}[h]
  \centering
    \includegraphics[width=\textwidth]{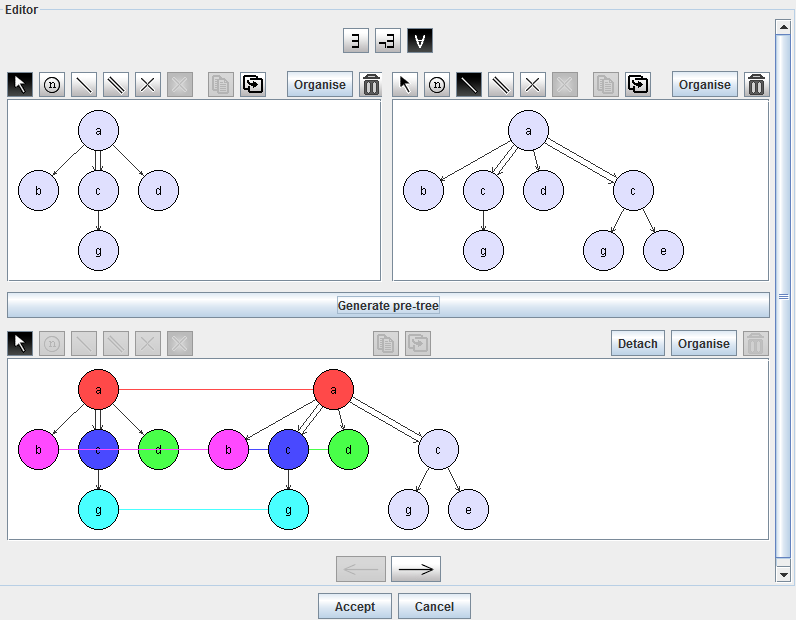}
  \caption{Editor for conditional constraints}
  \label{fig:grafico2}
\end{figure}

\subsection{Introducing Clauses}
The Home screen of the application consists of three different panels, besides the menu bar: the clauses' panel, the constraints' panel, and the pattern editor.
In order to create a clause, click on the ``Edit'' button of ``New Clause''. After that, the constraints' panel will show the selected clause's constraints. Since the clause is new, it will only appear the option of creating a new constraint. By clicking on the ``Edit'' button of ``New Constraint'', the pattern editor will be shown, as it can be noticed in Figure~\ref{fig:grafico1}.
The type of constraint ($\exists, \neg \exists, \forall$) is indicated by clicking on one of the upper buttons. Then, build the pattern by using the nodes, children edges (/), and descendant edges (//) creation buttons.

If the selected constraint is conditional, $\forall(c: p  \rightarrow q)$ , the editor screen will change into the one in Figure~\ref{fig:grafico2},
where $p$ must be drawn on the upper left box and $q$ on the upper right box. 
Once they are set, click on the ``Generate pre-tree'' button and the system will find all the prefix functions that exist between the two patterns. 
Click on the arrow-form buttons to choose the correct function and click on ``Accept''.

Once we have drawn all constraints of all clauses, we can save this specification by selecting the option ``Save'' or ``Save as'' from the ``File'' menu.
It is saved with a name and extension ``.spec''. 

It is also possible to load an existing specification (that was previously saved) by selecting the option ``Open'' from the ``File'' menu.

\subsection{Executing the Refutation Procedure}

\begin{figure}[h]
  \centering
    \includegraphics[width=0.5\textwidth]{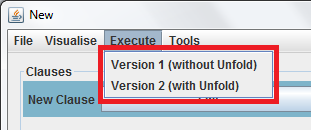}
  \caption{Menu for executing the procedure}
  \label{fig:grafico3}
\end{figure}

\begin{figure}[h]
  \centering
    \includegraphics[width=0.45\textwidth]{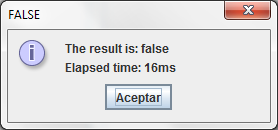}
  \caption{The result of the procedure}
  \label{fig:grafico4}
\end{figure}
\begin{figure}
  \centering
    \includegraphics[width=0.8\textwidth]{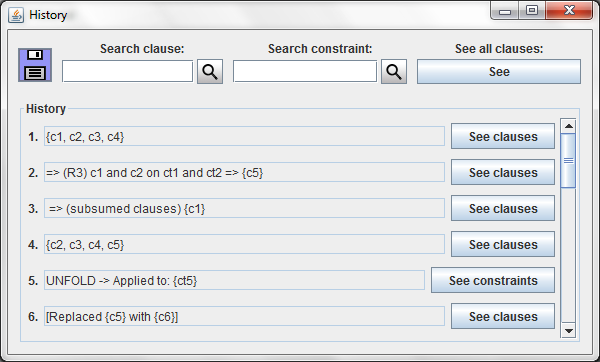}
  \caption{History of the procedure}
  \label{fig:grafico5}
\end{figure}

Once all the clauses have been created, let us see how to run the refutation procedure. To do so, pick one of the two versions from the ``Execute'' menu, as it is shown in Figure~\ref{fig:grafico3}.
As said before, when we choose to run ``Version 1'', the procedure consists of the three inference rules (R1), (R2), and (R3) together with some rules for deleting and simplifying clauses.
After finishing the procedure (if it stops), a message is displayed in which appears the satisfiability result of the input specification and the elapsed time (see Figure~\ref{fig:grafico4}).
If this result is \emph{FALSE} then we have proven that the specification is unsatisfiable. If it stops without obtaining \emph{FALSE}, then we cannot affirm that the specification is satisfiable because the procedure is not complete.
If we choose to run ``Version 2'', this procedure adds some rules to perform the previously mentioned ``unfolding process''.
By using this new procedure, we can prove unsatisfiability in more specifications than with the ``Version 1'' procedure, but for now it is an ongoing work.

The execution history will be automatically opened in other window. This history is very useful to see the rules used to obtain \emph{FALSE}. In this new screen (shown in Figure~\ref{fig:grafico5}), besides the history, it is possible to consult clauses and constraints. Enter the identifier of a clause (e.g. c4) in the clause searching area and it will be shown. The constraint search works exactly like the clause search, but with constraint identifiers (e.g. ct1). The upper right button loads every existing clause and displays them.
Also, with the ``See clauses'' buttons, it is possible to consult specific clauses from the different steps of the history. For instance, if we click on the button of the second step in Figure~\ref{fig:grafico5}, the system will load the clauses c1, c2, and c5.
Finally, to export the history to a text file, click on the save button on the upper left corner.

\subsection{Document Checking}

\begin{figure}[h]
  \centering
    \includegraphics[width=0.5\textwidth]{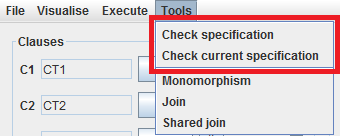}
  \caption{Document checking option}
  \label{fig:grafico6}
\end{figure}

Another important aim of this application is to check whether a given document satisfies a given specification or not. For that, click on ``Check specification'' in the ``Tools'' menu  (see Figure~\ref{fig:grafico6}).
This operation will open a window, similar to the Home screen, where the clauses of the specification and the XML document will be introduced.
The application also allows one to copy the set of clauses from the Home screen to this window. For that, click on ``Check current specification'' in the ``Tools'' menu (see Figure~\ref{fig:grafico6}). After being copied, new clauses can be introduced or existing ones can be deleted without compromising the original ones. In this case, the XML document must be introduced too.

\begin{figure}[h]
  \centering
    \includegraphics[width=\textwidth]{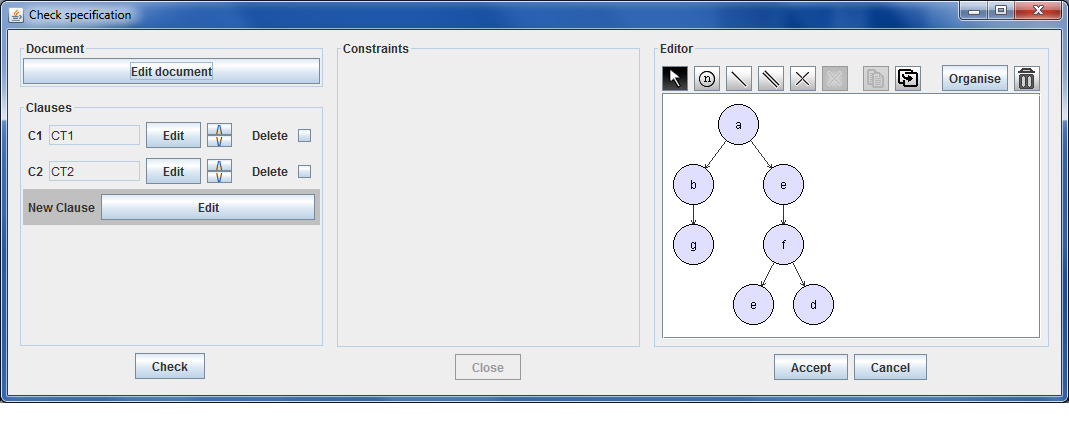}
  \caption{Document checking against a specification}
  \label{fig:graficoModel}
\end{figure}

\begin{figure}[h]
  \centering
    \includegraphics[width=\textwidth]{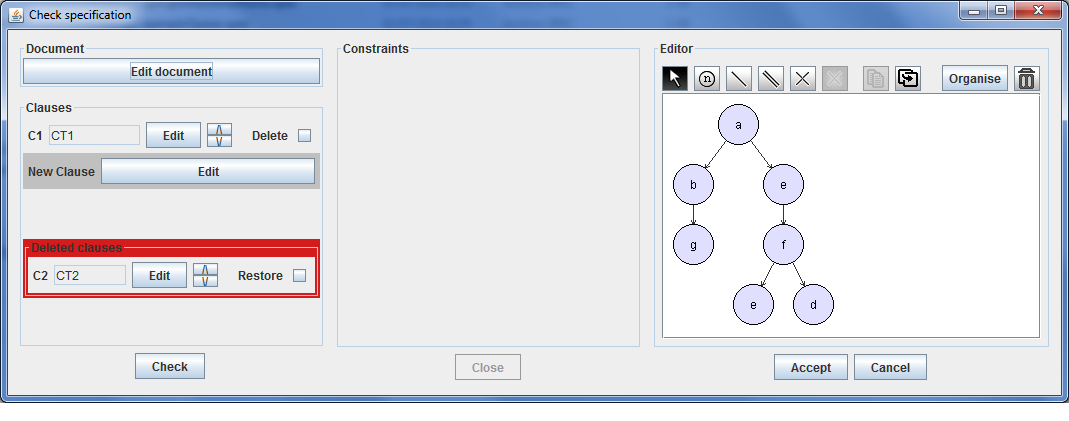}
  \caption{Document checking against some clauses in a specification}
  \label{fig:graficoModel2}
\end{figure}

After introducing the XML document by clicking on the ``Accept'' button, and once loaded the specification by any of the two possible ways,
 click on the ``Check'' button (see Figure~\ref{fig:graficoModel}) and a message with the result will be shown.
 The message will be \emph{TRUE} when the document is a model of the specification, and \emph{FALSE} otherwise.
 In the later case, we  probably want to check which are the clauses that the document satisfies and which are not. 
Clauses can be deleted temporarily from the specification by clicking on its ``Delete'' button to do these tests and they can be later restored by clicking on its ``Restore'' button (see Figure~\ref{fig:graficoModel2}).

\subsection{Other Tools}

Throughout the refutation process three basic operations are continuosly used: monomorphism from $p$ into $q$, the operation $p_1\otimes p_2$ in rule (R2), and the operation $p_1\otimes_{c,m}q$ in rule (R3). The application includes tools to execute such operations visually called {\em Monomorphism},  {\em Join}, and  {\em Shared join}, respectively.

\subsubsection{Monomorphism}

\begin{figure}[h]
  \centering
    \includegraphics[width=\textwidth]{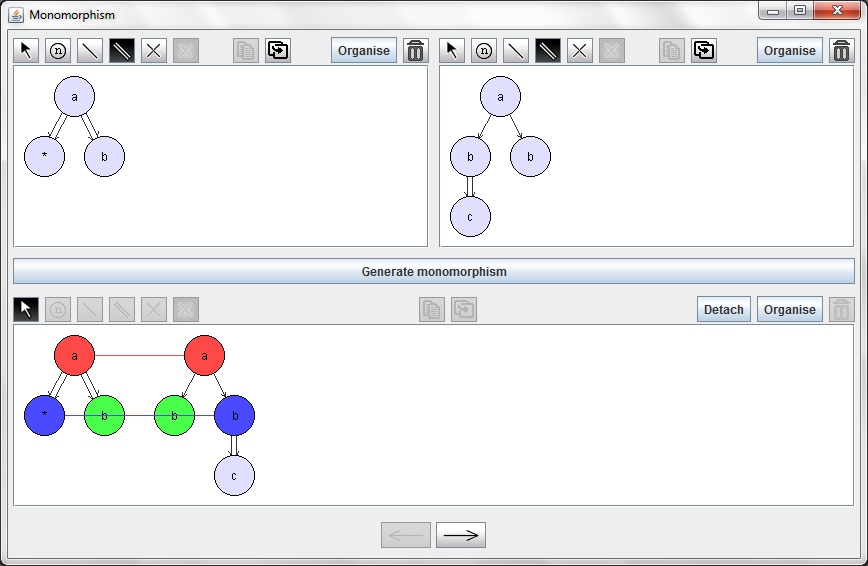}
  \caption{Monomorphism}
  \label{fig:grafico9}
\end{figure}

When selecting ``Monomorphism'' from the ``Tools'' menu, a new screen will appear, very similar to the one for creating a conditional constraint. Provided that we want to find out whether there exists a monomorphism from $p$ into $q$, we introduce the pattern $p$ into the upper left box and the pattern $q$ into the right box. Then, click on ``Generate monomorphism'' and the system will find every possible solution. For instance in the  Figure~\ref{fig:grafico9} it is shown one of the four monomorphisms existing from $p$ into $q$. The other three solutions can be consulted by clicking on the arrow-form buttons. 

\subsubsection{Join operation ($p_1\otimes p_2$)}

\begin{figure}
  \centering
    \includegraphics[width=\textwidth]{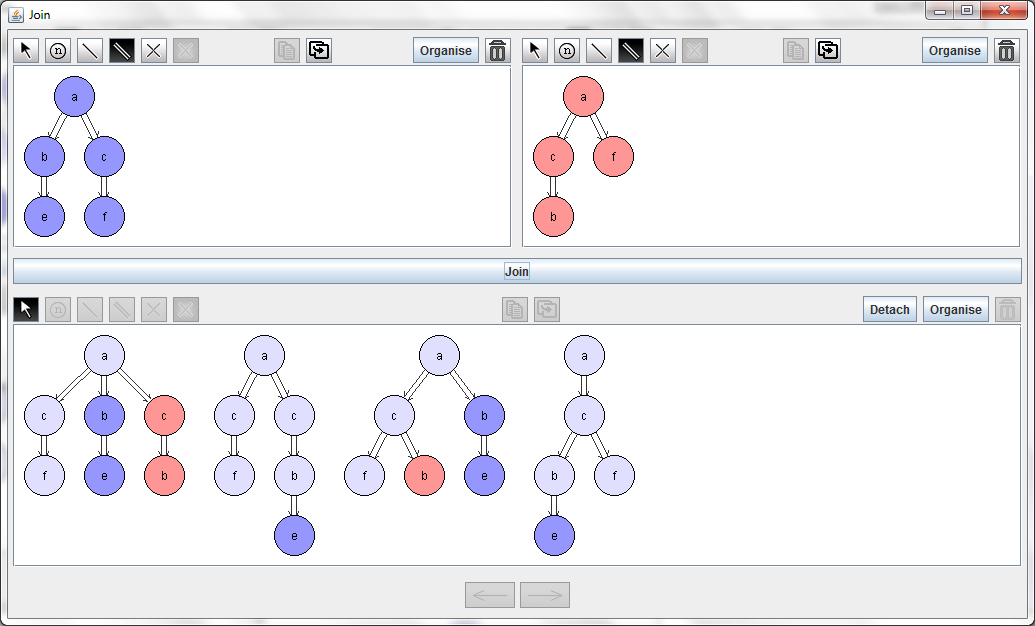}
  \caption {Join operation $p_1\otimes p_2$}
  \label{fig:grafico10}
\end{figure}

The ``Join'' tool will also open a similar window to the conditional constraint screen. We will introduce the patterns we want to operate, $p_1$ and $p_2$, into the two upper editors. After that, we click on the ``Join'' button and the solution will be calculated. On the lower editor will appear a set of patterns $s_{1}, s_{2}, ..., s_{n}$ which express the different ways of ``combining'' $p_1$ and $p_2$ (see Figure~\ref{fig:grafico10}). Recall this operation is used in rule (R2) to obtain  $(\bigvee_{s \in {p_1 \otimes p_2}} \exists  s)$ from $\exists p_1$ and $\exists p_2$.

\subsubsection{Shared join operation ($p_1\otimes_{c,m}q$)}

\begin{figure}
  \centering
    \includegraphics[width=\textwidth]{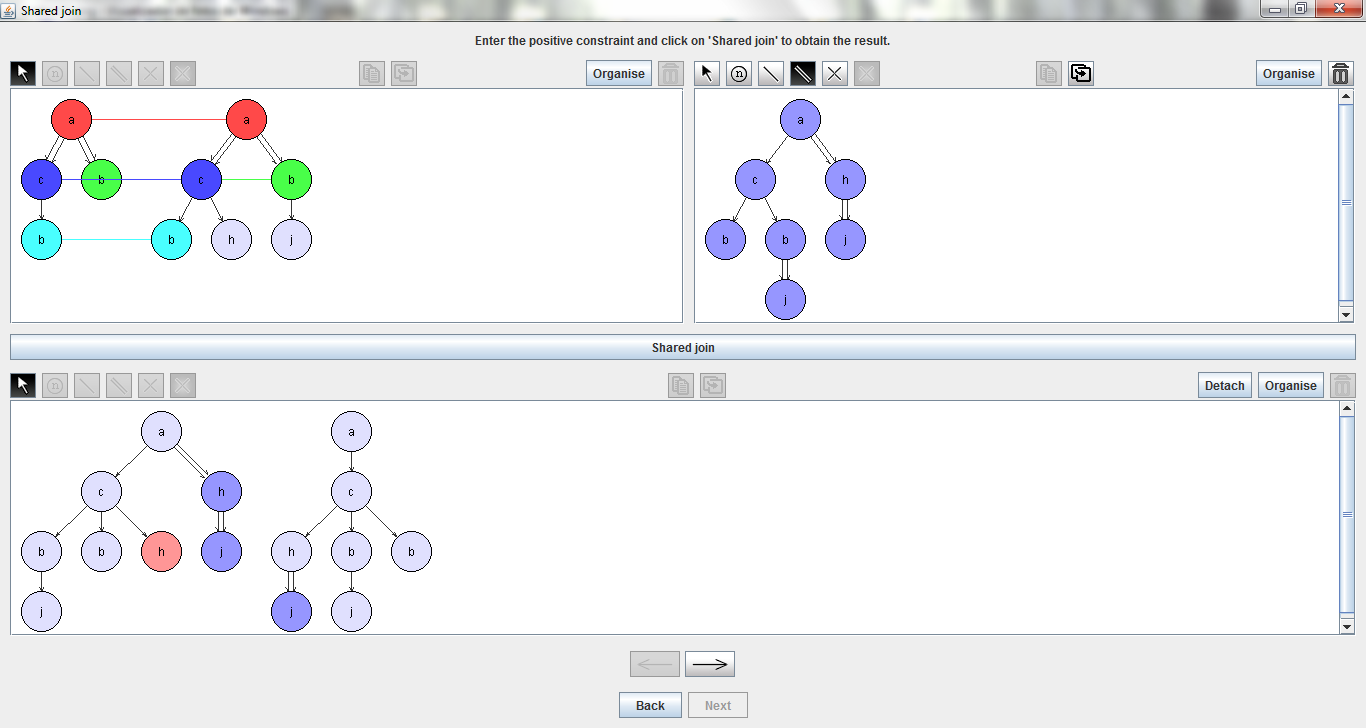}
  \caption{Shared join operation $p_1\otimes_{c,m}q$}
  \label{fig:grafico11}
\end{figure}

Similarly, the operation $p_1\otimes_{c,m}q$ is used in rule (R3) to obtain $(\bigvee_{s \in p_1 \otimes_{c,m} q} \exists  s)$ from 
the constraints $\exists  p_1$ and $\forall (c: p_2 \rightarrow q)$.
Since one of them is a conditional constraint, the tool is comprised by two windows. In the first one, we will introduce the conditional constraint as shown in Figure~\ref{fig:grafico2} and, after clicking on the ``Next'' button, this conditional constraint appears on the upper left box of the second window (see  Figure~\ref{fig:grafico11}); whereas  the positive constraint is introduced into the upper right box. Then, we click on ``Shared join'' and in the lower editor will appear a set of patterns $s_{1}, s_{2}, ..., s_{n}$ which express the different ways of ``combining'' $p_1$ and $q$ by considering the morphism $m$ from $p_2$ to $p_1$. Due to the possibility of having more than one monomorphism from $p_2$ to $p_1$ (that cannot be extended to a monomorphism from $q$ to $p_1$), different solutions will be shown. We can change the solution by clicking on the arrow-form buttons.

\section{Implementation Notes} \label{imp notes}

The prototype implementing the previously described refutation procedure is available at \url{http://www.sc.ehu.es/jiwnagom/PaginaWebLorea/SpecSatisfiabilityTool.html}, where we also explain the application's requirements and the configuration of the Java-Prolog bridge. The code of this application consists of around 1300 Prolog lines (in SWI-Prolog version 6.0.2) for the refutation procedure and around 4000 Java lines (in Java version jre7) for the interface.

Now, we roughly explain the algorithms designed in each version of the refutation procedure. See \cite{Albors2014} for more details about the implementation or for a user guide of the application.

\subsection{Version 1 algorithm}

We give here the idea of the algorithm  implementing this refutation procedure. We start with the initial specification $S_{0}$. Clause by clause and constraint by constraint the procedure applies every possible inference rule (R1), (R2), or (R3) obtaining a set $S'_{0}$ of new clauses.
Now the system divides the application of the rules in two parts: first, it applies every possible rule between two clauses, but being one from $S_{0}$ and the other one from the new set $S'_{0}$. After finishing this part, it applies every possible rule among the clauses in $S'_{0}$. In this way, all the clauses (resolvents) produced by applying these rules on 
$S_1$ = $S_0 \cup S'_0$ are in a new set $S'_{1}$.
This process will be repeated until the clause \emph{FALSE} comes out or until no rule can be applied. 

As said above, other rules for deleting and simplifying clauses are also applied (as soon as possible) in order to get a better performance. 
In particular, each time a new clause is produced by an inference rule, the procedure tries to simplify or delete this clause or another one related to this one. 
For instance,  a clause like $ \Gamma_1 \vee \Gamma_2 $ can be deleted from the set of clauses if the clause $ \Gamma_1 $ is produced as a resolvent (and therefore it is added to the set), since the latter one subsumes the first one. On the other hand, a clause containing two equal literals can be simplified by deleting one of them in the clause.  

By using all these rules, the inference rules  (R1), (R2), and (R3),  and the rules for deleting and simplifying clauses,  the input specification and each specification $S_i$ obtained during this process are logically equivalent. Therefore, given a specification as input, if the result of running ``Version 1'' is that the procedure stops with \emph{FALSE}, then the specification is unsatisfiable. Otherwise (if it does not stop or if it stops without obtaining \emph{FALSE}) then we cannot give any answer about the specification satisfiability.

\subsection{Version 2 algorithm}

This version is an ongoing work. The actual algorithm is as follows: It starts by calling to the Version 1. Then, if the result returns the clause \emph{FALSE}, it finishes (since it has been proven that the input specification is unsatisfiable). If Version 1 finishes returning \emph{TRUE}, then the ``unfolding process'' is done. If this process does not obtain new clauses, the algorithm finishes with the result of \emph{TRUE}, meaning that the input specification is satisfiable. But if the ``unfolding process'' obtains a new set of clauses, then the whole procedure (``version 1'' + ``unfolding process'') is repeated with this new set of clauses as input specification. 

Apart of trying to prove the completeness of the algorithm (as said previously), we are also testing the actual implementation. Some problems with non-termination due to rule (R3) and the fairness property must still be fixed (a procedure is fair whenever no inference rule is postponed forever). Nevertheless, for some specification examples, we can prove that they are unsatisfiable by running  ``version 2''   while they cannot be proven running ``version 1''.

\bibliographystyle{eptcs}
\bibliography{biblio}
\end{document}